**Effect of nucleation sites on the growth and quality of single-crystal boron arsenide**

Geethal Amila Gamage[1], Ke Chen[2], Gang Chen[2], Fei Tian[1*], Zhifeng Ren[1*]

[1] *Department of Physics and Texas Center for Superconductivity, University of Houston, Houston, Texas 77204, USA*
[2] *Department of Mechanical Engineering, Massachusetts Institute of Technology, Cambridge, MA 02139, USA*
*Corresponding authors. E-mail addresses: feifei131305@163.com (F. Tian), zren@uh.edu (Z. Ren).

**Abstract**

Boron arsenide (BAs) has been the least investigated cubic III-V compound, but it has recently attracted significant attention since the confirmation of its unusually high thermal conductivity above 1000 W m$^{-1}$ K$^{-1}$. However, determining how to achieve growth of a BAs single crystal on the centimeter scale remains unsolved, which strongly limits further research into, and potential applications of, this interesting material. Here we report our technique to grow a 7-mm-long BAs single crystal *via* the chemical vapor transport method by applying an additional nucleation site. The different thermal conductivity values obtained from BAs single crystals grown on nucleation sites of different compositions show the importance of choosing the proper nucleation-site material. We believe these findings will inspire further research into the growth of this unique semiconductor.

Advanced by the silicon (Si) industry, techniques to grow single crystals to certain dimensions and with specific electrical, optical, or thermal properties are both the basis for academic research and instrumental to the development of practical devices. Increasingly greater numbers of useful semiconducting materials, *e.g.*, silicon carbide (SiC)[1-3] and gallium arsenide (GaAs)[4], have benefited from the development of such production techniques and are now widely used in daily life. With the development of modern supercomputers, additional findings on novel and existing advanced materials from first-principles calculations are being investigated[5]. In 2013, L. Lindsay *et al.* reported an exception among such advanced materials: cubic boron arsenide (BAs)[6-7], which consists of a heavy element (As) but still exhibits the second highest room-temperature (RT) thermal conductivity ($\kappa$)[7-10] among bulk materials[11-14]. More surprisingly, in addition to the outstanding $\kappa$ of BAs and its unique behavior under pressure[15], its predicted remarkable high carrier mobility[16] and confirmed mild bandgap[17-19], as well as its stable mechanical performance[20] and proper thermal expansion coefficient[21-22], indicate that BAs is a promising semiconductor integrating both high power efficiency and fast heat dissipation.

However, further studies and applications of this unique material are limited by the difficulties in growing BAs large size single crystals (SCs)[23]. Inspired by T.L. Chu and A.E. Hyslop[24], the chemical vapor transport (CVT) method has been employed to grow BAs SCs. Multiple groups have begun to optimize CVT parameters for BAs growth, *i.e.*, precursors[25], transport agents[26], temperature[27], pressure[28], *etc.*, but the grown crystals have mostly been only around 1 mm long with lots of defects[29-30], far too small to be easily handled. Furthermore, determining how to control the amount and location of the nucleation remains unknown: BAs always randomly nucleates greatly on the inner wall of the quartz ($SiO_2$) tube, which results in dense growths that compete with one another. Here we report our findings that by inserting hetero materials into the CVT growth end to serve as additional nucleation sites, the nucleation process and the growth direction can both be well-controlled, such that a 7-mm-long BAs single crystal is finally obtained.

Details of the BAs CVT growth process have been fully described by Tian *et al.* in 2018[13]. High purity boron (B), arsenic (As), and iodine ($I_2$) are placed at the source end of a quartz tube, a piece of quartz bar is placed at the growth end, and the tube is sealed. As has been mentioned by many groups, BAs crystals grow on the inner wall of the quartz tube. Such uncontrolled nucleation is not preferred during the growth process, but it also suggests that quartz is a good nucleation site for BAs. Hence, we introduced a quartz bar into the tube to serve as an additional nucleation site and found that, using this technique, nucleation mainly occurred at the bar and a 7-mm-long BAs SC (Figure 1a) was obtained, the X-ray diffraction pattern for which is shown in Figure 1c. We also found that while many of the SCs grow at various angles to the bar, the largest one grows perpendicular to the bar (Figure 1b). Hence, if the bar sits parallel to the tube-length direction, BAs grows within the cross-sectional plane of the tube, and the final size of the SC is limited by the tube's diameter; on the other hand, BAs grows along the tube-length direction if the bar is placed perpendicular to the tube-length direction, and the final size of the SC is limited by the tube's length. For this experiment, we used 6-mm-diameter quartz tubes, so when the quartz bar was placed parallel to the tube's length direction, the collected crystals were always less than 6 mm in diameter, while the 7-mm-long BAs SC (Figure 1a) was grown on a bar placed perpendicular to tube's length direction. Since the tube's length is always much larger than its diameter, placing the quartz bar in a certain position at the growth end will help control the nucleation and allow the growth of a large BAs SC along the preferred direction.

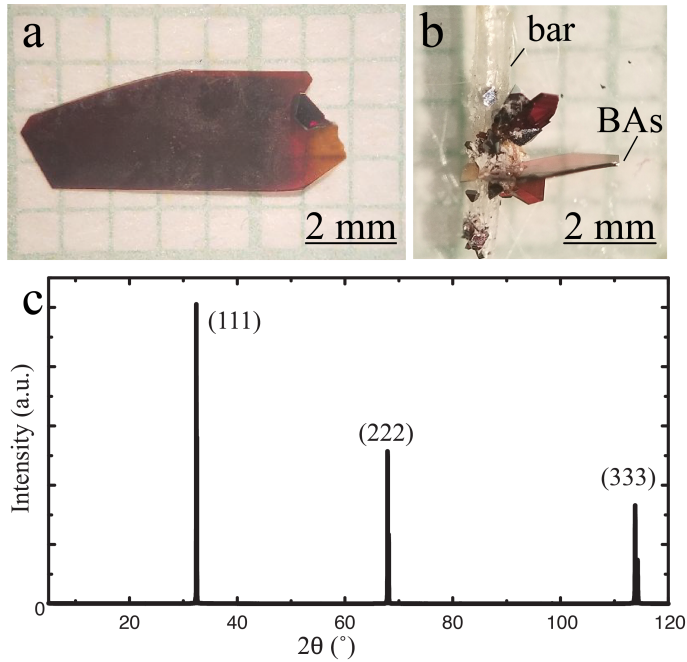

**Figure 1**. Photographs of (a) the obtained 7-mm-long BAs SC and (b) growth of BAs SCs on a quartz bar. (c) X-ray diffraction (Cu Kα source) pattern of the 7-mm-long BAs SC shown in (a).

At the same time, growing a BAs SC in this way will not affect the quality of the crystal. We used the time domain thermoreflectance (TDTR) method to characterize its $\kappa$ at RT, similar to previous reports[11-13]. An 80-nm-thick aluminum transducer layer was deposited onto the BAs sample surface by an e-beam evaporator. A laser pulse train with 800 nm central wavelength and 80 MHz repetition rate was generated from a Ti:sapphire laser oscillator. A beam splitter was used to divide the beam into pump and probe beams. The intensity of the pump laser was modulated sinusoidally by an electro-optic modulator (EOM) at 3 MHz, focused on the metal surface with a 30 μm diameter ($1/e^2$) and 45 mW power, and acted as a heating source that generated a temperature change at the metal surface. The probe laser pulse was delayed by a motorized translation stage, focused and overlapped with the pump spot at the metal surface with a 10 μm diameter ($1/e^2$) and 10 mW power. Thermoreflectance of the probe beam as a function of the delay time was recorded by a silicon photodiode, the output of which was connected to a lock-in amplifier (LIA) working at the EOM frequency. The phase (PHI) of the LIA as a function of the delay time, which is determined by thermal transport in the material system, was taken as the measured signal. A three-dimensional Fourier heat conduction model for the Al-BAs structure was built to fit the measured data and to extract the thermal conductivity of the BAs substrate along with the interface thermal conductance between the metal layer and the BAs sample. As shown in Figure 2a, the fitting curve agrees well with the measured phase signal, yielding a RT $\kappa$ of 960 ± 90 W m$^{-1}$ K$^{-1}$, quite close to the reported values when no additional nucleation sites are introduced[11-13]. The simulated curves using thermal conductivity 20% larger or smaller than the fitted one are well separated away from the measured signal, showing the excellent sensitivity of the measurement. As shown in Figure 2b, the uncertainty of the fitted thermal conductivity was estimated by the standard deviation of the Gaussian distribution of a simulation of 600 TDTR measurements using the Monte Carlo method[31], which considers both the experimental noise and the propagation of the error of the known parameters in the model.

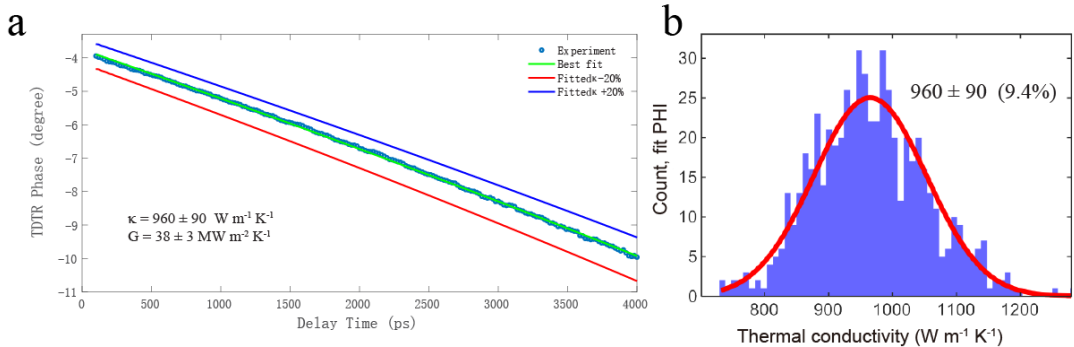

**Figure 2**. (a) TDTR phase signal of the BAs sample, along with its best fitting curve and simulated curves using thermal conductivities 20% larger and smaller than the fitted value. (b) Distribution of the fitted thermal conductivities from a Monte Carlo simulation of 600 TDTR measurements[31].

To further investigate the influence of the nucleation site on the growth and also thermal conductivity, high-quality GaAs, Si, and sapphire (Al$_2$O$_3$) bars were each chosen to serve as the additional nucleation site for BAs SC growth, rather than the quartz bar. After the same 2-week growth time, several-mm-long BAs SCs were collected from the GaAs bar, while on the Si and sapphire bars, only some small pieces of BAs SC were found. The maximum RT $\kappa$ values obtained from the BAs SCs grown from different nucleation sites are shown in Figure 3. The RT $\kappa$ values of BAs SCs grown on quartz (1240 ± 100 W m$^{-1}$ K$^{-1}$)[32] and GaAs (1240 ± 110 W m$^{-1}$ K$^{-1}$) bars are close to the predicted value, while the maximum $\kappa$ values of the BAs crystals grown on sapphire and Si bars are much lower.

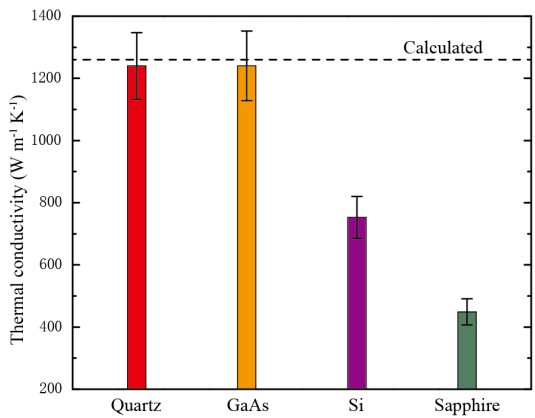

**Figure 3**. Maximum RT $\kappa$ data measured by the TDTR method from BAs SCs grown using different nucleation sites. Uncertainty for the thermal conductivity values was estimated by using the Monte Carlo method[31]. The dashed line represents the calculated RT $\kappa$ for BAs considering 3-phonon, 4-phonon, and phonon-isotope scattering[13].

Since the $\kappa$ (especially ultrahigh $\kappa$) value is highly sensitive to various defects, the $\kappa$ data clearly show the effects of different nucleation sites and on the quality of each grown BAs SC. In the previously reported BAs SC growth techniques in which high $\kappa$ values were obtained[11-13], the growth environment initially includes four compositions: B, As, I$_2$, and quartz. It should be noted that a small amount of Si impurity coming from the quartz tube[29], as well as commonly occurring defects (*i.e.*, twin boundaries[13], vacancies[25], and antisite pairs[30]), are understood to be responsible for the failure to grow uniform BAs SCs. In this case, when a quartz bar is selected as the additional nucleation site, there are no other kinds of material introduced, and the quality of the final product will not be affected, which can be proven by the observed near-record-high $\kappa$ value. Additionally, GaAs is quite stable and does not react in this environment below 900 °C, so a GaAs bar works as well as a quartz bar as the nucleation site. However, when sapphire is used as the nucleation site, it reacts with $I_2$, which disturbs the reactions in the CVT process and suppresses the high $\kappa$. A similar situation occurs when Si is used (Si + $2I_2$ = $SiI_4$). Hence, both quartz and GaAs nucleation sites lead to high-quality BAs SCs, while Si and sapphire nucleation sites result in low $\kappa$.

To summarize, we have grown a 7-mm-long BAs SC by a nucleation-site-introduced CVT method. A quartz or GaAs bar, serving as an additional nucleation site, is placed at the growth end and helps control the nucleation position and growth direction. By confirming the near-record-high RT $\kappa$ value of the obtained crystal by TDTR, we conclude that the introduction of the quartz or GaAs bar does not reduce the intrinsic ultrahigh $\kappa$. We believe our findings will provide insight into determining how to grow larger and better BAs single crystals and will help establish the foundation for studying and designing BAs-based devices.

## Acknowledgement

This work is supported by the Office of Naval Research under MURI grant N00014-16-1-2436.